\documentclass{webofc} 
  \usepackage[varg]{txfonts}   
\usepackage{graphicx}
\usepackage{amsmath}
\usepackage[utf8]{inputenc}
\usepackage{color}
\newcommand{\TeV}{\ \mathrm{TeV}}
\newcommand{\GeV}{\ \mathrm{GeV}}

\newcommand{\dd}{ {\mathrm d} } 

\newlength{\szovszel}
\newlength{\slashszel}

\begin{document}
\title{Mass hierarchy and energy scaling of the Tsallis\,--\,Pareto pa\-ra\-me\-ters in hadron productions at RHIC and LHC energies}
\author{\firstname{G\'abor} \lastname{B\'ir\'o}\inst{1,2}\fnsep\thanks{\email{biro.gabor@wigner.mta.hu}} \and
        \firstname{Gergely G\'abor} \lastname{Barnaf\"oldi}\inst{1}\and
        \firstname{Tam\'as S\'andor} \lastname{Bir\'o}\inst{1}\and
        \firstname{Keming} \lastname{Shen}\inst{3}
}
\institute{Wigner Research Centre for Physics of the H.A.S., P.O. Box, H-1525 Budapest, Hungary
\and
          Institute of Physics, E\"otv\"os University, 1/A P\'azm\'any P. S\'et\'any, H-1117 Budapest, Hungary
\and
          Key Laboratory of Quark \& Lepton Physics and Institute of Particle Physics, Central China Normal University, Wuhan 430079, China
}
%
\abstract{
The latest, high-accuracy identified hadron spectra measurements in high-energy nuclear collisions led us to the investigation of the strongly interacting particles and collective effects in small systems. Since microscopical processes result in a statistical Tsallis\,--\,Pareto distribution, the fit parameters $q$ and $T$ are well suited for identifying system size scalings and initial conditions. Moreover, parameter values provide information on the deviation from the extensive, Boltzmann\,--\,Gibbs statistics in finite-volumes.
We apply here the fit procedure developed in our earlier study for proton-proton collisions~\cite{BG:entr17, BG:maxent16}. The observed mass and $\sqrt{s}$ energy trends in the hadron production are compared to RHIC dAu and LHC pPb data in different centrality/multiplicity classes. Here we present new results on mass hierarchy in pp and pA from light to heavy hadrons.  
}
%
\maketitle
\section{Introduction}
\label{sec:intro}
In high-energy nuclear physics the investigation of small colliding nuclear systems is fundamentally important for understanding collective effects. Strong correlation phenomena typical for heavy-ion collisions were recently observed in pp and pA collisions~\cite{ALICE:nature, ALBERTA:collecticity}. While correlations in such small systems are more clearly observable thanks to the lower background, their statistical and thermodynamical description points towards to exceed the classical Gibbs\,--\,Boltzmann framework. 
%
%
A modern tool for such studies is provided by the non-extensive statistics. In the application one fits Tsallis\,--\,Pareto-type distributions to describe the measured spectra in wide central-of-mass energy and $p_T$ ranges~\cite{TS:nonext88, TSB:tphysica13, CL:jphys17, WILK:EPJ15}. In our recent works~\cite{BG:entr17, BG:maxent16}, we analyzed the spectra of various hadron species in pp collisions from RHIC to LHC energies. We compared the results to theoretical model calculations and we explored the mass hierarchy and c.m. energy scaling of the parameters. 
In this short contribution, we extend our investigations to the latest identified strange- and heavy-flavored hadron spectra measured in dAu and pPb collisions from $\sqrt{s_{NN}}=200\GeV$ RHIC to $\sqrt{s_{NN}} = 5.02\TeV$ LHC energies in different centralities. We used data from STAR~\cite{STAR:dAu} and ALICE~\cite{ALICE:pPb1, ALICE:pPb2} measurements and analyzed them following the procedure described in details in Refs.~\cite{BG:entr17, BG:maxent16}.
%
\section{Identified hadron spectra from pA collisions, compared to pp}
\label{sec:spectra}
Recent high-precision measurements in dAu and pPb collisions at RHIC and LHC energies present a {\sl Janus-face} behavior, depending on the multiplicity class of the selected data set. While low-multiplicity events are more jet-like, the high multiplicity ones include correlations and present sizeable nuclear modification, especially at the low- and intermediate momentum ranges. The same duality has been already observed in nucleus-nucleus (AA) collisions~\cite{ORTIZBENCEDI:JPG17, SOFTHARD:JPCS15, LILIN:AHP17}. In smaller systems it is a special challenge to identify these components~\cite{CL:NUCLEFF16}.
We quantify and separate these observations and deduce the parameters of the microscopical interactions. We apply the same formula as in Refs.~\cite{BG:entr17, BG:maxent16}, for all centrality classes (0-20\%, 20-40\% and 40-100\% for the STAR data, 0-5\%, 5-10\%, 10-20\%, 20-40\%, 40-60\%, 60-80\% and 80-100\% for the ALICE data):
\begin{equation}
\left.\frac{\dd^2N}{N_{ev}2 \pi p_T \dd p_T \dd y}\right|_{y\approx0}\sim \left[1+(q-1)\frac{m_T}{T}\right]^{-\frac{1}{q-1}} \ \ \ .
\label{eq:tsp}
\end{equation}
Here, $m_T=\sqrt{p_T^2+m^2}$ is the {\sl transverse mass} of the given hadron species. We note that this formula can be derived from the Tsallis $q$-entropy and, as we pointed out in Ref.~\cite{TSB:EPJ12, TSB:tphysica13}, the parameters carry a certain physical meaning: the {\sl non-extensivity} parameter, $q$ measures the deviation from the (extensive) Boltzmann\,--\,Gibbs state ($q=1$), while a generalized temperature-like parameter, $T$ provides a temperature value taking into account the finite-size effects and fluctuations. These parameters are interconnected via the multiplicity distribution, e.g. a (negative) binomial distribution~\cite{TSB:SQM16}.
%
\begin{figure}[htp]
  \vspace*{-0.2cm}
\begin{center}
\includegraphics[width=0.31\textwidth]{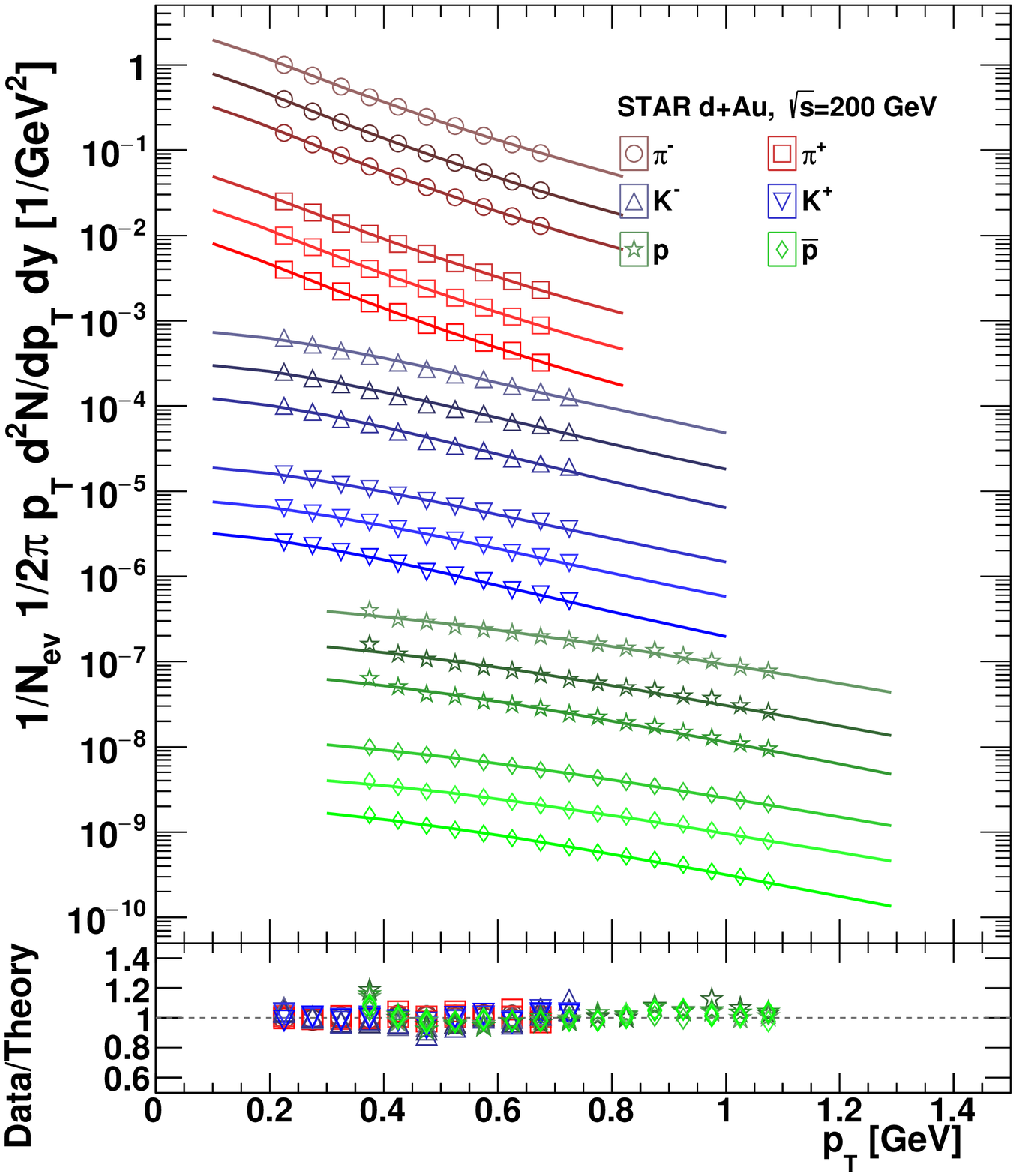}
\includegraphics[width=0.31\textwidth]{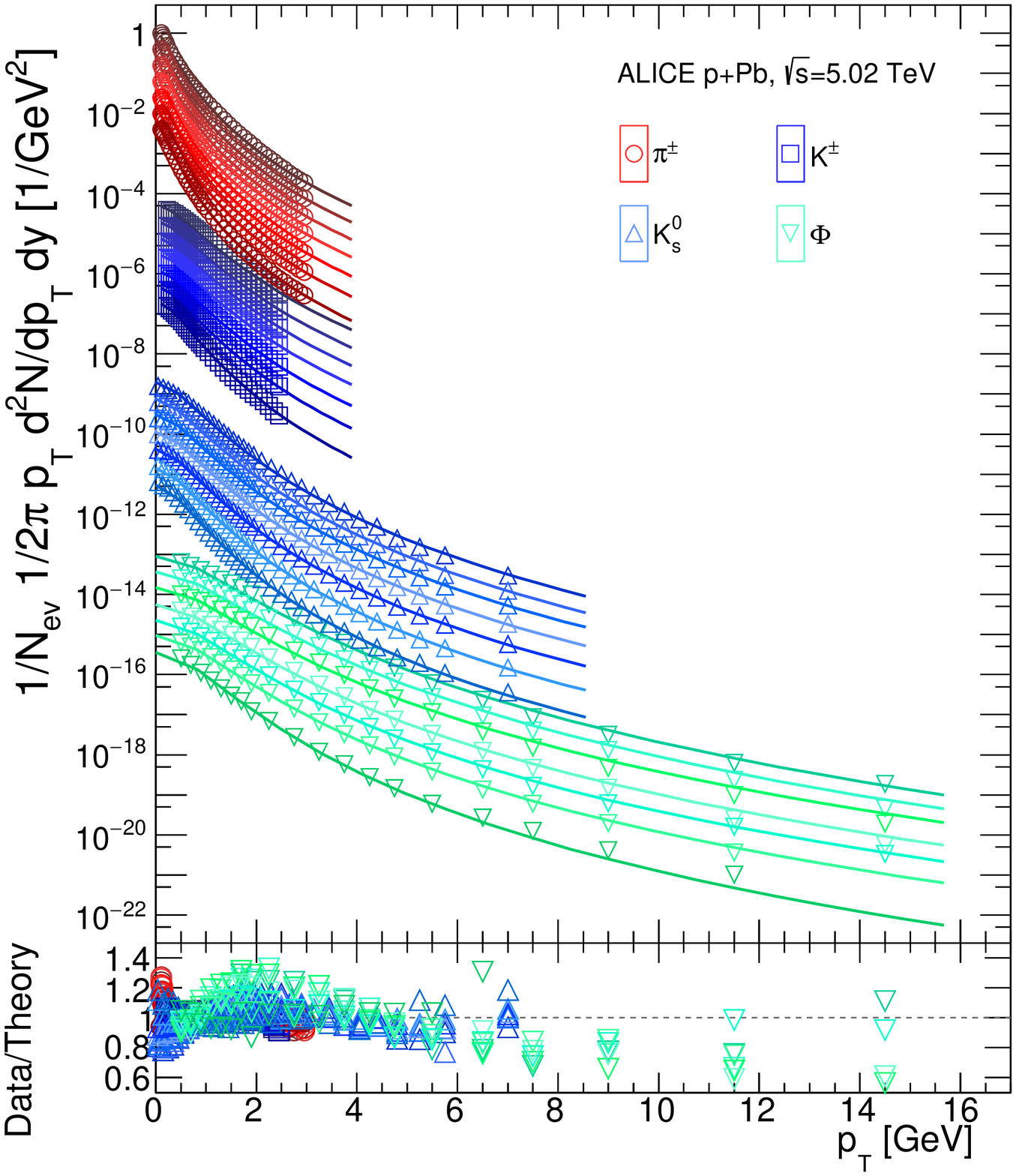}
\includegraphics[width=0.31\textwidth]{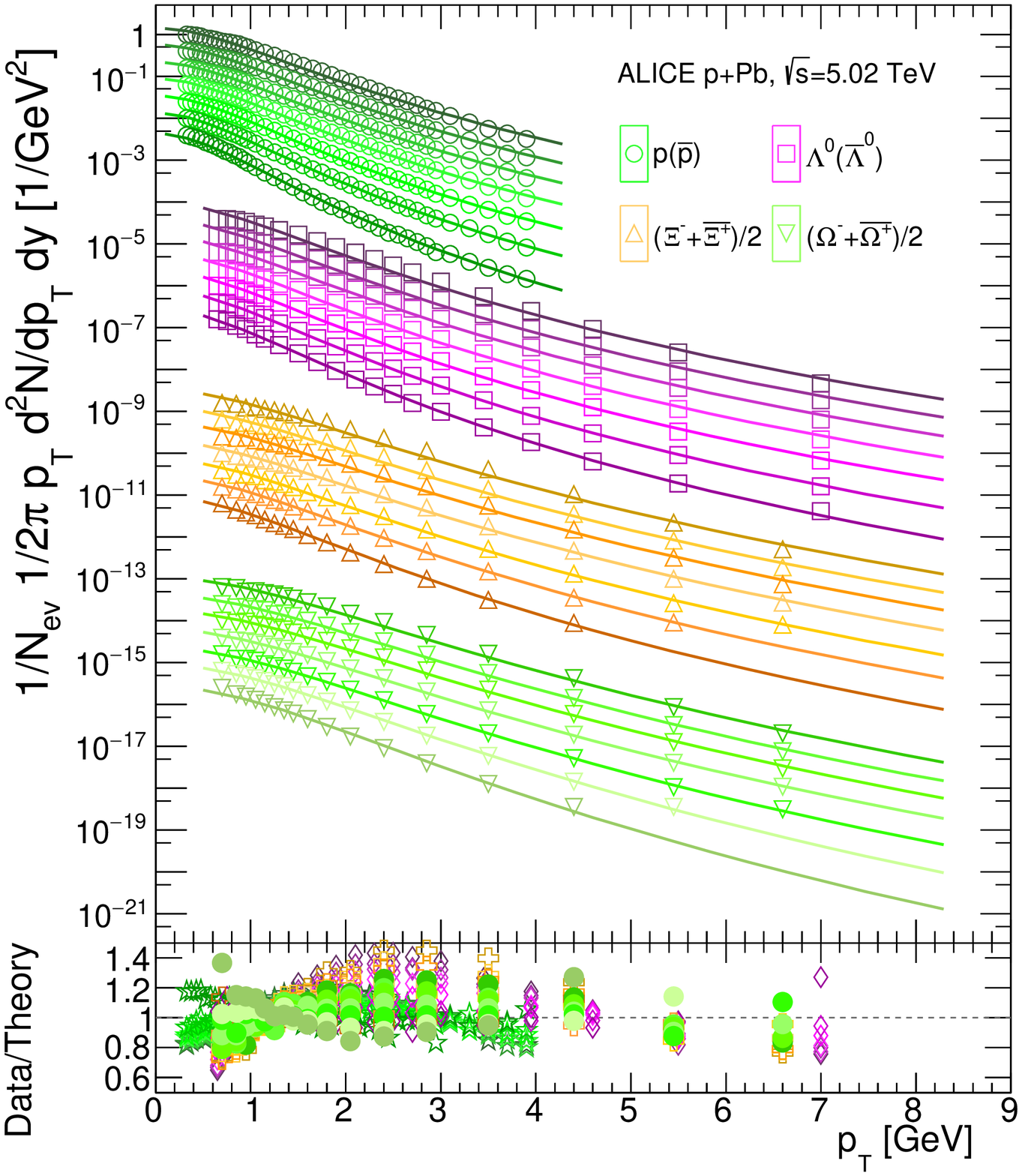}
\end{center}
\caption{\label{fig:ppb_sectra} The fitted spectra for various identified hadron spectra measured in dAu collisions at $\sqrt{s_{NN}}=200\GeV$ at STAR, RHIC ({\sl left}) and pPb collisions at $\sqrt{s_{NN}}=5.02\TeV$ at ALICE, LHC ({\sl middle panel for mesons and right panel for baryons}). Data have been scaled for better visibility. {\sl Bottom panels} present the Data/Fit ratios using the same color as on the spectra.}
\vspace*{-0.5cm}
\end{figure}
On Fig.~\ref{fig:ppb_sectra} the analyzed identified hadron spectra measured in small colliding systems along with the fitted distributions were plotted. For both the $\sqrt{s_{NN}}=200\GeV$ dAu at RHIC and $\sqrt{s_{NN}}=5.02\TeV$ pPb at LHC energy collisions we extraced the parameters of Eq.~\eqref{eq:tsp} utilizing the ROOT data analysis framework, performing fits based on the minimal $\chi^2$ method. Based on the available identified hadron data published in Refs.~\cite{STAR:dAu,ALICE:pPb1,ALICE:pPb2}, the fits were performed in various centrality bins on as wide $p_T$ ranges as possible. Lower panels present the data-to-fit ratio with agreement of $20\%$ and $40\%-50\%$ for dAu (RHIC) and pPb (LHC) data respectively. 
%
\section{Results: parameters and mass hierarchy}
\label{sec:results}
\begin{figure}[htp]
  \begin{center}
  \includegraphics[width=0.54\textwidth]{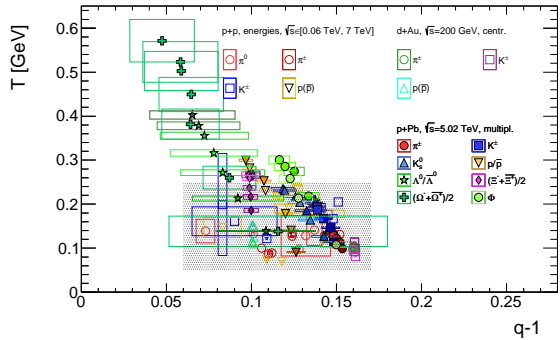}
  \end{center}
  \vspace*{-0.6cm}
  \caption{\label{fig:par_qT} The $T$ vs $(q-1)$ parameter space for identified hadron spectra in pPb at $\sqrt{s_{NN}}=5.02\TeV$ with different multiplicity classes ({\sl filled markers}) and for the same pp measurements for various $\sqrt{s}$ energies from Ref.~\cite{BG:entr17}  including dAu STAR data~\cite{STAR:dAu} at $\sqrt{s}=200$ GeV ({\sl empty markers and shaded area}).}
  \vspace*{-0.4cm}
  \end{figure}

\begin{figure}[htp]
  \vspace*{-0.5cm}
  \begin{center}
  \includegraphics[width=0.48\textwidth]{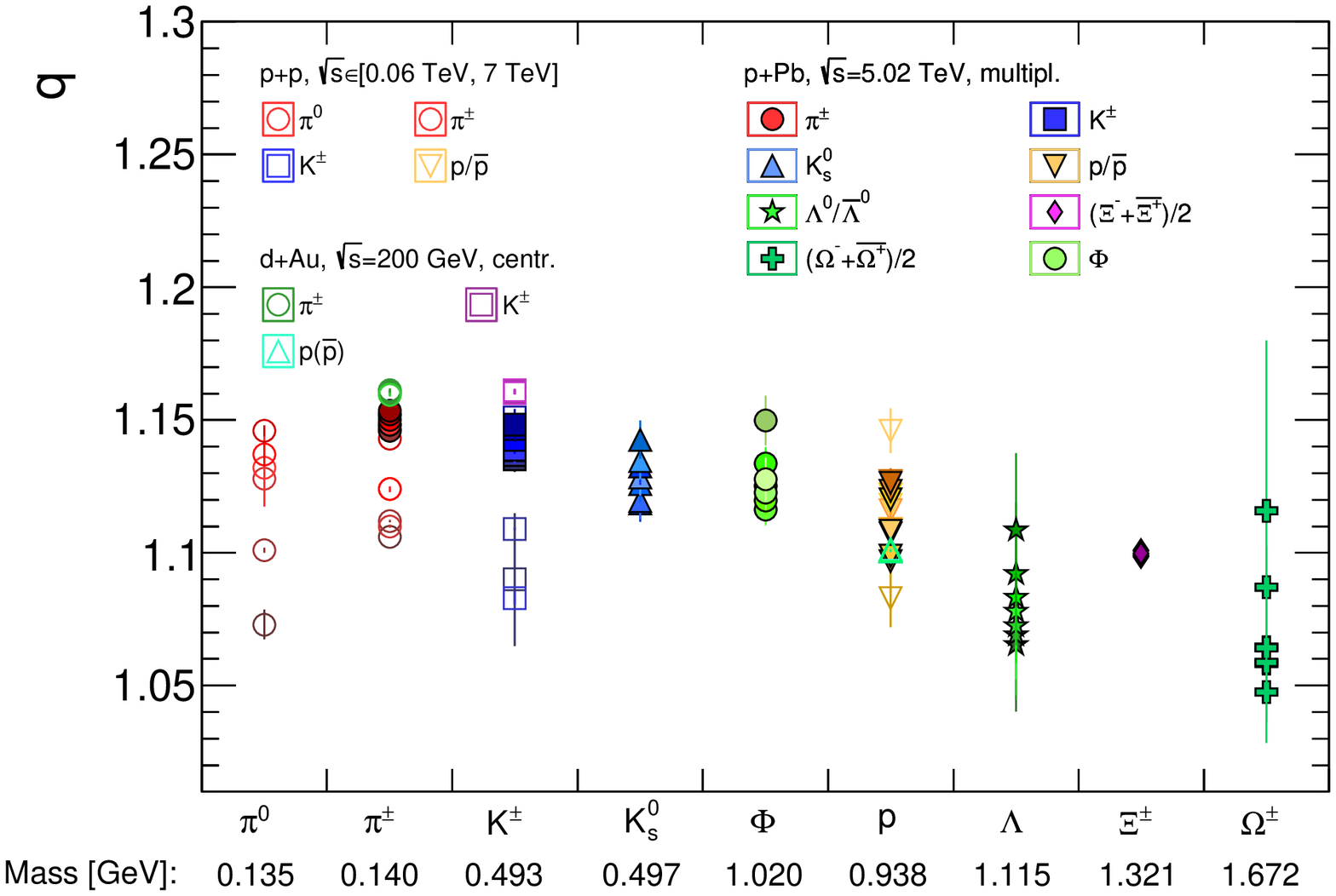}
  \includegraphics[width=0.48\textwidth]{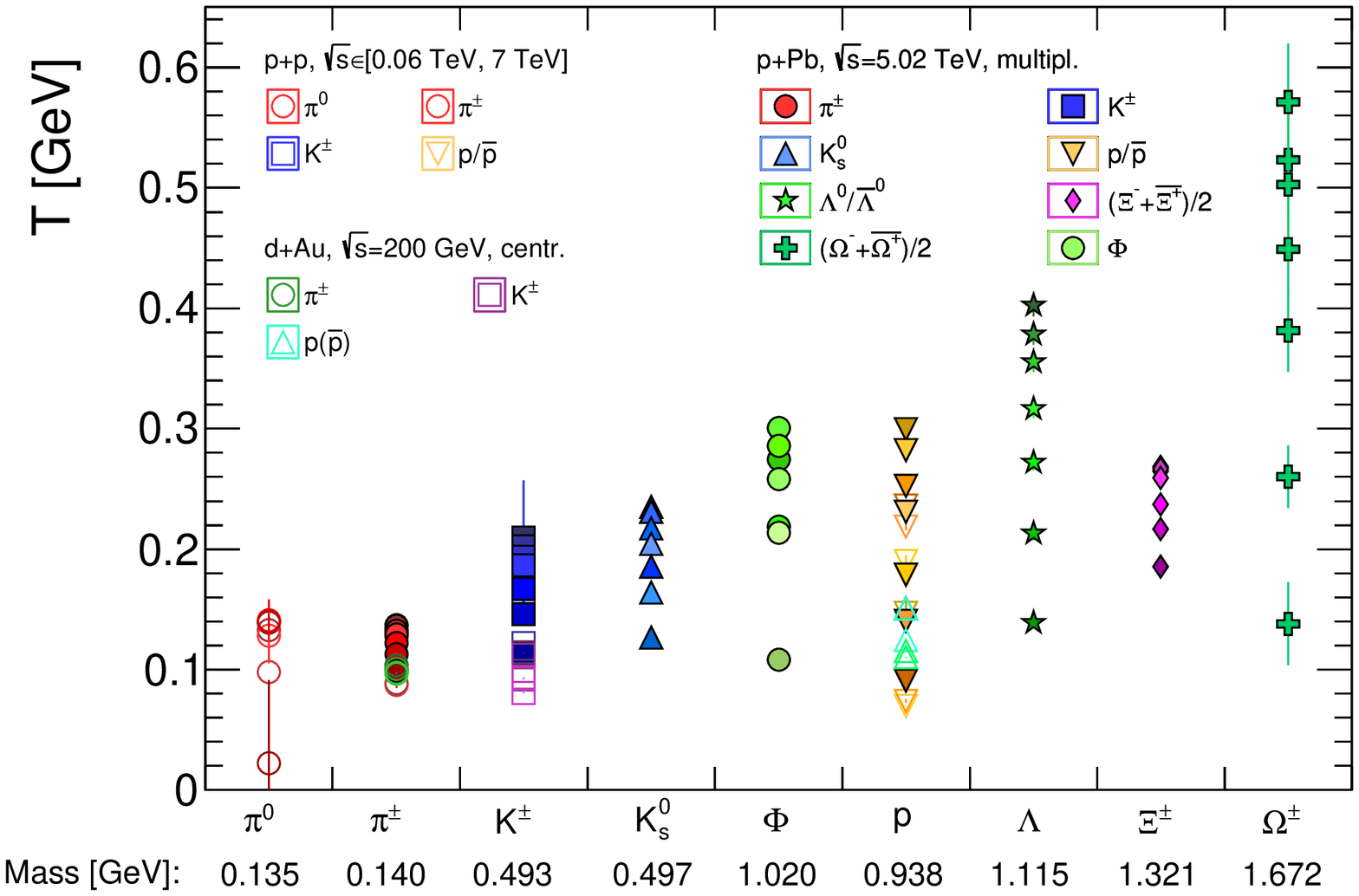}
\end{center}
\vspace*{-0.4cm}
  \caption{\label{fig:par_mass} The fitted Tsallis\,--\,Pareto parameters $q$ ({\sl left panel}) and $T$ ({\sl right panel}) obtained in pp, dAu, and pPb collisions for all investigated hadron species in increasing mass order.}
  \vspace*{-0.4cm}  
  \end{figure}

The fit parameters of Eq.~\eqref{eq:tsp} are plotted on the parameter-space diagram, Fig.~\ref{fig:par_qT}. $T$ is plotted agains $q-1$, the latter measures the deviation from the Boltzmannian case directly. The empty markers and the shaded area show the pp results with different energies ($62.4\GeV<\sqrt{s}<7\TeV$) as it was presented in Ref.~\cite{BG:entr17} along with the STAR dAu data at $\sqrt{s}=200\GeV$. The full markers show the parameters from the present fit for the pPb spectra at $\sqrt{s_{NN}} = 5.02\TeV$. 
The new points that are consistent with the previous results (i.e. those inside the shaded box) are the most pp-like points, especially for the smallest multiplicity class. We observe deviance only for those points which are extracted from high-multiplicity spectra.
Increasing the multiplicity either by increasing the c.m. energy or getting to more central events, lead to both higher $q$ and $T$. Points in the parameter space originating form datasets with similar multiplicities group together nicely.
The dAu data from STAR accompany the lowest energy pp fit parameters. It is either due to the low multiplicity or due to the very limited $p_T$ range of the identified transverse momentum spectra. The absence of strong power-law tail results in a weak determination of the Tsallis\,--\,Pareto parameters.
%
%
Fig.~\ref{fig:par_qT} also represents the mass hierarchy: the higher the mass of the identified hadron, the higher the $T$ and the lower the $q$ is. This effect is more visible on Fig.~\ref{fig:par_mass}, where the $q$ ({\sl left panel}) and $T$ ({\sl right panel}) parameters are plotted for each identified hardon species in order to analyze the mass hierarchy effects. Here we used the same markers as on the figures above. This result extends our earlier analysis in pp collisions in Refs.~\cite{BG:entr17, BG:maxent16} using only $\pi$, $K$, and $p(\bar{p})$, while adding the hadron species with higher masses, we can see a mass hierarchy for both parameters.
Based on the extracted parameters, the hadron species with the highest mass has the highest $T$, meanwhile the non-extensitivity is reduced as $q \to 1$, especially for the low multiplicity events. Trends with growing {\sl s} quark content supports the idea of strangeness enhancement~\cite{ALICE:nature}. It can be the signature of the quark-gluon plasma like, close-to-thermalized, hot and dense matter in small colliding nuclear systems like pp, dAu, or pPb.  
%
%
%
%
\section{Summary}
\label{sec:summary}
Mass and strangeness hierarchy is presented and analyzed for pp, dAu, and pPb collisions. We analyzed data in a wide $\sqrt{s_{NN}}$ range form RHIC to LHC energies in various centrality/multiplicity bins. The results are consistent with our previous center-of-mass energy scaling investigations for the most pp-like, peripheral centrality classes, however deviations are observed for central (c.f. high multiplicity) collisions for hadron species with high mass. 
Our main observation is that the measure of the non-extensitivity, $q-1$, decreases with increasing hadron mass, thus higher-mass hadrons present less deviation form the Boltzmann\,--\,Gibbs $q=1$ case. On the other hand, similarly to the exponential fits, higher-mass hadron species required higher $T$ to obtain the best fit results.
%
\section{Acknowledge}
\label{sec:ack}
This work was supported by the Hungarian-Chinese cooperation grant No TéT 12 CN-1-2012-0016 and No. MOST 2014DFG02050, Hungarian National Research Fund (OTKA) grants K120660 and K123815. 
Author G.G. Barnaf\"oldi also thanks the THOR COST action CA15213. We acknowledge the support of the Wigner GPU Laboratory.
%

%

\begin{thebibliography}{}
%
\bibitem{BG:entr17}
G.~Biro {\it et al.}, Entropy {\bf 88}, 19(3) (2017) 
%
\bibitem{BG:maxent16}
G.~Biro {\it et al.}, AIP Conf. Proc. {\bf 1853} 080001 (2017)
%
\bibitem{ALICE:nature}
ALICE Collaboration, Nat. Phys. {\bf 13}, 9 (2017)
%
\bibitem{ALBERTA:collecticity}
A. Toia, J. Phys. Conf. Ser. {\bf 798} 012068 (2017)
%
\bibitem{TS:nonext88}
C.~Tsallis, J. Stat. Phys. {\bf 52}, 479 (1988)
%
\bibitem{TSB:tphysica13}
Bir\'o, T.S.; V\'an, P.; Barnaf\"oldi, G.G.; V\'an, P. Eur. Phys. J. {\bf A49}, 110 (2013)
%
\bibitem{CL:jphys17}
J. Cleymans, J. Phys. Conf. Ser. {\bf 779}, 012079 (2017)
%
\bibitem{WILK:EPJ15}
M. Rybczynski; G. Wilk; Z. W\l{}odarczyk, EPJ Web Conf. {\bf 90} 01002 (2015)
%
\bibitem{GR:physrev17}
S. Grigoryan, Phys.Rev. {\bf D95}, 056021  (2017)
%
\bibitem{PAR:eurph17}
A.S. Parvan, O.V. Teryaev, J. Cleymans, Eur.Phys.J. {\bf A53}, 102 (2017)
%
\bibitem{STAR:dAu}
STAR Collaboration, Phys. Rev. C. {\bf 79}, 034909 (2009)
%
\bibitem{ALICE:pPb1}
ALICE Collaboration, Phys.Lett. B. {\bf 728}, 25-38 (2014)
%
\bibitem{ALICE:pPb2}
ALICE Collaboration, Phys.Lett. B. {\bf 758}, 389-401 (2016)
%
\bibitem{ORTIZBENCEDI:JPG17}
A. Ortiz; G. Benc\'edi; H. Bello, J. Phys. G: Nucl. Part. Phys. {\bf 44}, 065001 (2017)
%
\bibitem{SOFTHARD:JPCS15}
G. G. Barnaf\"oldi; K. \"Urm\"ossy; G. B\'ir\'o, J. Phys. Conf. Ser. {\bf 612} 012048 (2015)
%
\bibitem{LILIN:AHP17}
X. Yin; L. Zhu; H. Zheng, Adv. High Energy Phys. {\bf 2017} 6708581 (2017)
%
\bibitem{CL:NUCLEFF16}
S. Tripathy; T. Bhattacharyya; P. Garg; P. Kumar; R. Sahoo; J. Cleymans, Eur. Phys. J. A {\bf 52}, 289 (2016)
%
\bibitem{TSB:EPJ12}
T. S. Bir\'o; G. G. Barnaf\"oldi; P. V\'an, EPJ A {\bf 49}, 110 (2013)
%
\bibitem{TSB:SQM16}
T. S. Bir\'o; G. G. Barnaf\"oldi; G. B\'ir\'o; K.M. Shen, J. Phys. Conf. Ser. {\bf 779}, 012081 (2017)
%
\end{thebibliography}
\end{document}